\documentclass[printer]{aa}  

\usepackage{graphicx}
\usepackage{txfonts}
\usepackage{multirow}
\usepackage{bigdelim}
\begin{document}

   \title{Detection of a repeated transit signature in the light curve of the enigma star KIC\,8462852: a 928-day period?}

   \author{F. Kiefer\inst{1}\fnmsep\thanks{ \email{flavien.kiefer@iap.fr}}
          \and
          A. Lecavelier des \'Etangs\inst{1}
          \and
          A. Vidal-Madjar\inst{1}
          \and 
          G. Hébrard\inst{1,2}
          \and
          V. Bourrier\inst{3}
          \and
          P.A. Wilson\inst{1,4}
          }

   \institute{Institut d'Astrophysique de Paris, UMR7095 CNRS, Universit\'e Pierre \& Marie Curie,
98bis boulevard Arago, 75014 Paris, France            
         \and
             Observatoire de Haute-Provence, CNRS, Université d’Aix-Marseille, 04870 Saint-Michel-l’Observatoire, France        
         \and
             Observatoire de Genève, Chemin des Maillettes 51, Sauverny, CH-1290 Versoix, Suisse
         \and
             Leiden Observatory, Leiden University, Postbus 9513, 2300 RA Leiden, The Netherlands
             }

   \date{Received ; accepted }

  \abstract{

As revealed by its peculiar Kepler light curve, the enigmatic star KIC\,8462852 undergoes short and deep flux dimmings at \textit{a priori} unrelated epochs. It presents nonetheless all other characteristics of a quiet 1\,Gyr old F3V star. These dimmings resemble the absorption features expected for the transit of dust cometary tails. The exocomet scenario is therefore most commonly advocated. We reanalyzed the Kepler data and extracted a new high-quality light curve to allow for the search of shallow signature of single or a few exocomets. We discovered that among the 22 flux dimming events that we identified, two events present a striking similarity. These events occurred 928.25\,days apart, lasted for 4.4\,days with a drop of the star brightness by 1000\,ppm.
We show that the light curve of these events is well explained by the occultation of the star by a giant ring system, or the transit of a string of half a dozen of exocomets with a typical dust production rate of 10$^5$-10$^6$\,kg/s. Assuming that these two similar events are related to the transit of the same object, we derive a period of 928.25\,days. The following transit was expected in March 2017 but bad weather prohibited us to detect it from ground-based spectroscopy. We predict that the next event will occur from the 3$^{\rm rd}$ to the 8$^{\rm th}$ of October 2019. } 
 

   \maketitle
%

\section{Introduction}

KIC\,8462852 is a peculiar and intriguing source that caught a lot of attention from the astronomic community in the recent years \citep{Boyajian2016, Montet2016, Lisse2015, Wright2016, Bodman2016, Harp2016, Schaefer2016, Makarov2016, Schuetz2016, Thompson2016, Abeysekara2016, Marengo2015, Hippke2016, Neslusan2017}. KIC\,8462852 is an F3V star located at about 454$\pm$35\,pc away from us \citep{Hippke2016}. The Kepler Spacecraft photometric data revealed an enigmatic lightcurve for this star, with erratic, up to $\sim$20\% deep, stellar flux dimmings \citep{Boyajian2016}. Being otherwise considered a standard F-star, stellar unstabilities could be excluded to explain its strange behavior. More recently, it was found that KIC\,8462852's flux dropped by $\sim$2.5\% over 200\,days \citep{Montet2016}, while it was suspected from a thorough analysis of photographic plates taken over the last century to continuously decrease by about 0.3\%\,yr$^{-1}$ \citep{Schaefer2016}. 

The most popular scenario advocated to explain the frequent but aperiodic dips is that of many uncorrelated circumstellar objects transiting at different epochs; either comets \citep{Boyajian2016,Neslusan2017} or planetesimal fragments \citep{Bodman2016}. This is reminiscent of the case of $\beta$\,Pictoris, on the spectra of which many variable narrow absorptions were observed at high-resolution in Ca\,II doublet, best explained by extrasolar comets, or exocomets~\citep{Ferlet1987,Beust1990,Kiefer2014a}.  It should be noted however that contrary to $\beta$\,Pictoris, KIC\,8462852 is not young (1\,Gyr), and any circumstellar gas or dust remain unobserved at infrared wavelengths. 

About 20 years ago, \cite{Lecavelier1999a} published innovative simulations of photometric signatures produced by the transit of the dusty tail of exocomets. The shape of the theoretical absorption signatures obtained has some unique specificities: a peaky core for the transit of the head of the coma, and a long trailing slope. Nonetheless, the only direct evidence for comets around stars other than the Sun came from high-resolution spectroscopy, observing the atomic gas counterpart of the cometary tail, with \textit{e.g.} $\beta$\,Pic~\citep{Ferlet1987}, HD172555~\citep{Kiefer2014b}, HR10~\citep{Lagrange1990} or 49\,Ceti~\citep{Montgomery2012, Miles2016}. Before KIC\,8462852, photometry never revealed any direct observations of exocomets around any star.

The level of precision needed to detect the transit of a single $\beta$\,Pic like exocomet is about several 100\,ppm~\citep{Lecavelier1999a}. Detecting such object is a difficult task, since a single solar-like exocomet cannot be expected to transit several times during the lifetime of Kepler (in the Solar System, the comets have period typically larger than 3 years). Nevertheless, the opportunity of detecting repeated transits should not be completely excluded.
We selected KIC\,8462852 for thorough analysis of its Kepler lightcurve with the goal of finding single object 100\,ppm-deep transit signatures. We report in the present paper the detection of a 1000\,ppm deep signature repeating twice at 928\,days interval in KIC\,8462852 lightcurve. We successfully modeled this signature by a string of exocomets crossing the line of sight one after the other at 0.3\,AU from the central star. 

Alternatively, we found that at least one other scenario could provide a good fit of the lightcurve: the transit of a wide ring system surrounding a planet orbiting at 2.1\,au from the star. Hill-sphere could indeed become much wider than the star itself at distances larger than 1\,au, and contain transiting materials such as rings \citep{Kenworthy2015, Lecavelier2017, Aizawa2017}. Moreover, while plausible and simplistic, such scenario was recently proposed by~\cite{Ballesteros2017} to explain the smooth and solitary D800 dip of~\cite{Boyajian2016}, with a transiting ring planet on a 12 years orbit around the star. Thus, instead of being a comets host, KIC\,8462852 might just be a planetary system with at least two ring planets.

Kepler data reduction is presented in Section 2. The two identitical 1000\,ppm deep events are presented in Section 3. In Section 4 we show that these events are real and not instrumental or due to background objects. The modelisation of these events are presented in Section 5. Finally, we present in Sect. 6, an attempt to observe that event in March 2017 which failed due to bad weather, and a prediction on its future realisations in October 2019 and later.

\section{The Kepler photometric data reduction}
\label{sec:kepler_data}
\begin{figure*}\centering
\includegraphics[width=250mm, clip=true, trim=100 0 -300 0]{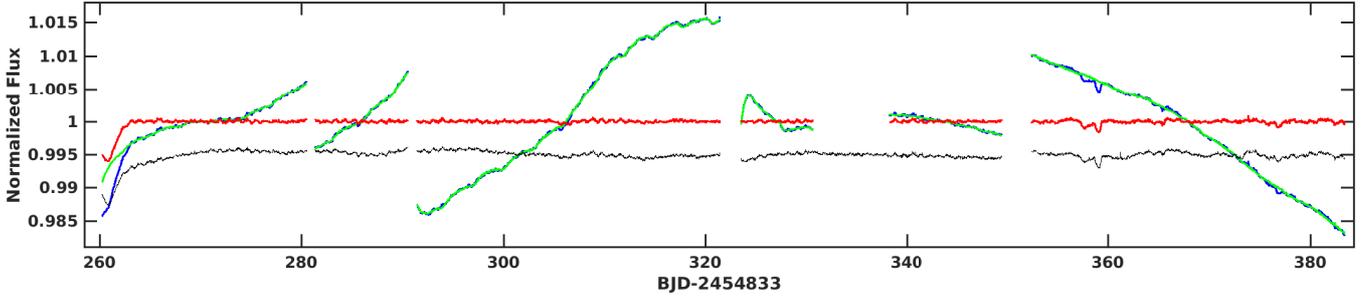}
\caption{\label{fig:CBV_fit1} Example of CBVs fit on lightcurves with small or no dips. In blue, the raw SAP lightcurve, in green the fitted CBVs, as explained in the text, and in red the final detrended SAP lightcurve. For comparison, we superimposed in black the PDCSAP data, with an offset of -0.005 for visual convenience. Green and blue curves overlap most of the time, but it can be seen on some shallow dips (\textit{e.g.} at BJD-2454833=360) that the fitted CBVs, in green, stay at the baseline level long these events.}
\end{figure*}
\begin{figure*}\centering
\includegraphics[width=250mm, clip=true, trim=100 0 -300 0]{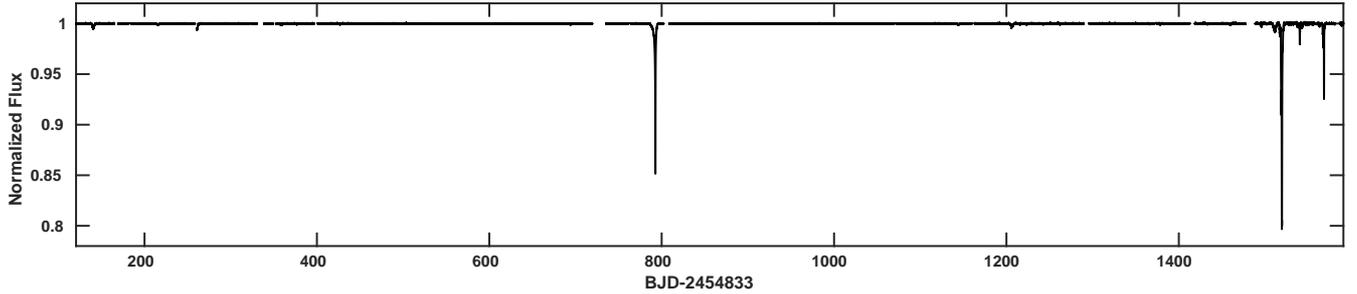}
\caption{\label{fig:full_LC} Full KIC\,8462852 detrended lightcurve.}
\end{figure*}

The Kepler spatial observatory \citep{Borucki2010} followed KIC\,8462852 in long cadence mode (30-min sampling) during about 4 years from the 2nd of May, 2009 to the 11th of May, 2013, separated into 17 quarters of continuous integration. The Kepler pipeline produced raw (simple aperture photometry, SAP) and reduced (pre-search data conditioning, PDC-SAP) lightcurves of the full 4 years time range \citep{Smith2012}. The SAP data essentially consist in calibrated flux but uncorrected of cosmic ray absorption, systematic behaviours, jumps etc. The PDCSAP lightcurve are systematically corrected for every trend of non-astrophysical origin by the reduction pipeline. While the PDCSAP data are certainly good enough for detecting short 0.1-1\% deep transits, they do not reach the level of precision needed to detect 0.1\% deep, possibly day-long, absorption signatures that could be typically produced by transiting exocomets~\citep{Lecavelier1999a,Kiefer2014a}. We thus wrote our own MATLAB-routine to carefully reduce the SAP lightcurve, which principles are explained below.

The Kepler pipeline determines for each quarter, in each CCD-channel, an ensemble of 16 Cotrending Basis Vector (CBV) calculated from the lightcurve of the brightest stars in the channel using Principal Component Analysis (PCA) \citep{Kinemuchi2012,Smith2012}. These CBVs represent the main systematic behaviours common in the lightcurves collected on the same channel during the same quarter. For each quarter, we fitted to KIC\,8462852's SAP lightcurve a linear combination of the 13 first CBVs. Following the recommended process \citep{Kinemuchi2012}, we increased iteratively the number of fitted CBVs. We stopped when the resulting lightcurve baseline was the flattest in the quiet periods, and that adding more CBVs led to no significant improvement. In order to avoid fitting out the physical dips, we iteratively excluded from the fit any data points below the continuum minus 2$\sigma$, with $\sigma$ defined as
\begin{equation}
\sigma = 1.48\times\text{MAD(data-continuum)}
\end{equation}

The median absolute deviation (MAD) times 1.48 is an estimation of the standard deviation, not biased by outliers. Jumps and spikes (cosmics etc.) were carefully filtered out before applying the fit. This is done by studying the first-derivative of the light curve and identifying spikes and jumps signatures. They appear as single or P-cygni shaped 3-5 cadences-long peaks, with an amplitude at least 4-times larger than the local typical cadence-to-cadence variations. None-to-ten measurements are found in such discontinuities per quarter. If a spike is encountered, the bad cadences are first removed and then the light curve is linearly interpolated through the resultant gap. If a jump is encountered, the bad cadences are removed and the lightcurve separated in two pieces around the gap; in this case, the CBVs are fitted out to each piece separatively.  

More generally, anytime there is missing data (typically more than 25 adjacent cadences) we separated the lightcurve in two pieces around the gap and fitted out CBVs independently for these two pieces. Most of these large discontinuities are due to monthly Earth downlink and usually followed by thermal relaxation \citep{Kinemuchi2012}. Even though most of the time the CBVs capture such variation, simple fitting ignoring cadences within the gap was not accurate enough. Separating the lightcurve around these discontinuities led to better results.

Several examples of the detrending results are displayed on Figs.~\ref{fig:CBV_fit1}. We compare them to the pipeline automatic PDC reduction, which in general presents quarter-long low amplitude variations along the curve, and especially around strong dips. With our reduction the continuum is flat, allowing the shallower dips to emerge more evidently than in the PDCSAP data. This shows the positive effect of excluding the dips measurements when fitting out the systematics. The full detrended lightcurve is plotted on Fig.~\ref{fig:full_LC}.


\section{Two identical photometric shallow events}
\label{sec:2events}

Using the light curve obtained in the previous section, we can identify the photometric events 
that happened during the four years of observation. 
The observed photometric variations shows two different patterns: there are (1) periodic-like variations 
and (2) short-time decreases of the star brightness.
With a period close to 1 day, the periodic modulation corresponds to the 0.88-day signal due to stellar rotation 
and already noticed by \cite{Boyajian2016}. Beyond these variations, the stars shows significant short-time
and sporadic variations, all of them are dips of the star brightness below the mean brightness observed during the quiet period. 
 
We screen the entire light curve and identified a total of twenty-two significant dips. 
Apart from the strong dips already listed by \cite{Boyajian2016}, we found several shallower dips,
some of them also identified by \cite{Makarov2016}. Table~\ref{tab:events} summarizes these detections. 

\begin{table*}\centering
\caption{\label{tab:events} List of detected photometric dips in KIC\,8462852 lightcurve. Events 2 and 13 are renamed respectively A and B in the rest of the paper.}
\begin{tabular}{lcccll}
\hline 
Event & Epoch         & Depth                            & Width          & Comment                  & Previous publications$^\dagger$\\
      & (JD-2454833) &  ($\log_{10}\Delta F/F$)  & (day) \\
\hline
  1    &    140.7    &  -2.30    &   1.70    &  cometary tail shape (triangular)     &  B2016, MG2016 \\
  2 / A   &    215.8    &  -2.96    &   1.98    &  similar to event 13     &  MG2016: modulation of PSF centroid   \\
  3    &    261.0    &  -2.23    &   1.19    &  cut by a gap            &  B2016, MG2016 \\
  4    &    357.9    &  -3.03    &   0.70    &  cometary tail shape (triangular)    &         \\
  5    &    359.0    &  -2.82    &   0.41    &  superimposed on event 4 & B2016\\
  6    &    376.6    &  -3.08    &   1.25    &  noisy surrounding       &  MG2016 \\
  7    &    427.1    &  -3.10    &   1.78    &  partly fitted by CBVs   &  B2016, MG2016 \\
  8    &    688.6    &  -3.18    &   0.68    &   \rdelim\}{4}{10pt}[~~series of 4 small dips]  &         \\
  9    &    694.3    &  -3.03    &   1.18    &  & \\
 10    &    700.6    &  -3.32    &   1.72    &  & \\
 11    &    706.7    &  -3.37    &   0.94    &  & \\
 12    &    792.6    &  -0.90    &   0.77    &  very deep event & B2016, MG2016\\
 13 / B   &   1144.1    &  -2.97    &   1.98    &  similar to event 2     & MG2016: instrumental jitter  \\
 14    &   1206.2    &  -2.38    &   2.54    &  narrow event upon a wide event & B2016, MG2016  \\
 15    &   1224.0    &  -3.02    &   1.63    &  shallow event& \\
 16    &   1496.0    &  -2.60    &   0.55    &  cometary tail shape  (triangular) & B2016\\
 17    &   1511.4    &  -2.10    &   2.24    &  preceding a much deeper event & \\
 18    &   1519.4    &  -0.71    &   1.00    &  very deep event       & B2016, MG2016 \\
 19    &   1540.4    &  -1.75    &   0.43    &  deep event            & B2016, MG2016 \\
 20    &   1542.9    &  -2.43    &   0.73    &  triangular              & \\
 21    &   1563.7    &  -2.60    &   0.89    &  cometary tail shape  (triangular) & \\
 22    &   1568.2    &  -1.21    &   1.03    &  deep event & B2016, MG2016 \\
\hline
\end{tabular} \\
\flushleft
$^\dagger$ MG2016 = \cite{Makarov2016}, Table 1 \\
$^{\hphantom{\dagger}}$ B2016 = \cite{Boyajian2016}, Table 1
\end{table*}

\begin{figure}\centering
\includegraphics[width=100mm, clip=true, trim=0 0 0 0]{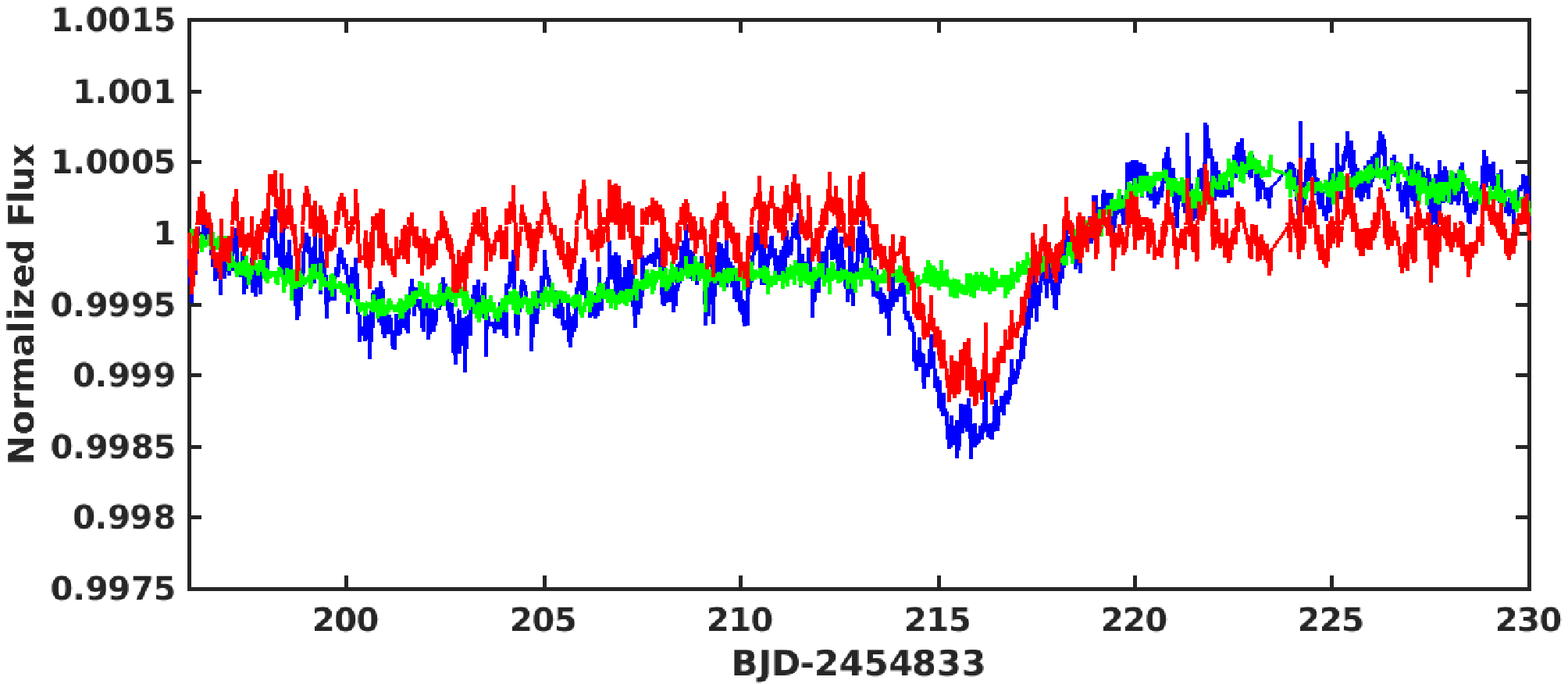}
\includegraphics[width=100mm, clip=true, trim=0 0 0 0]{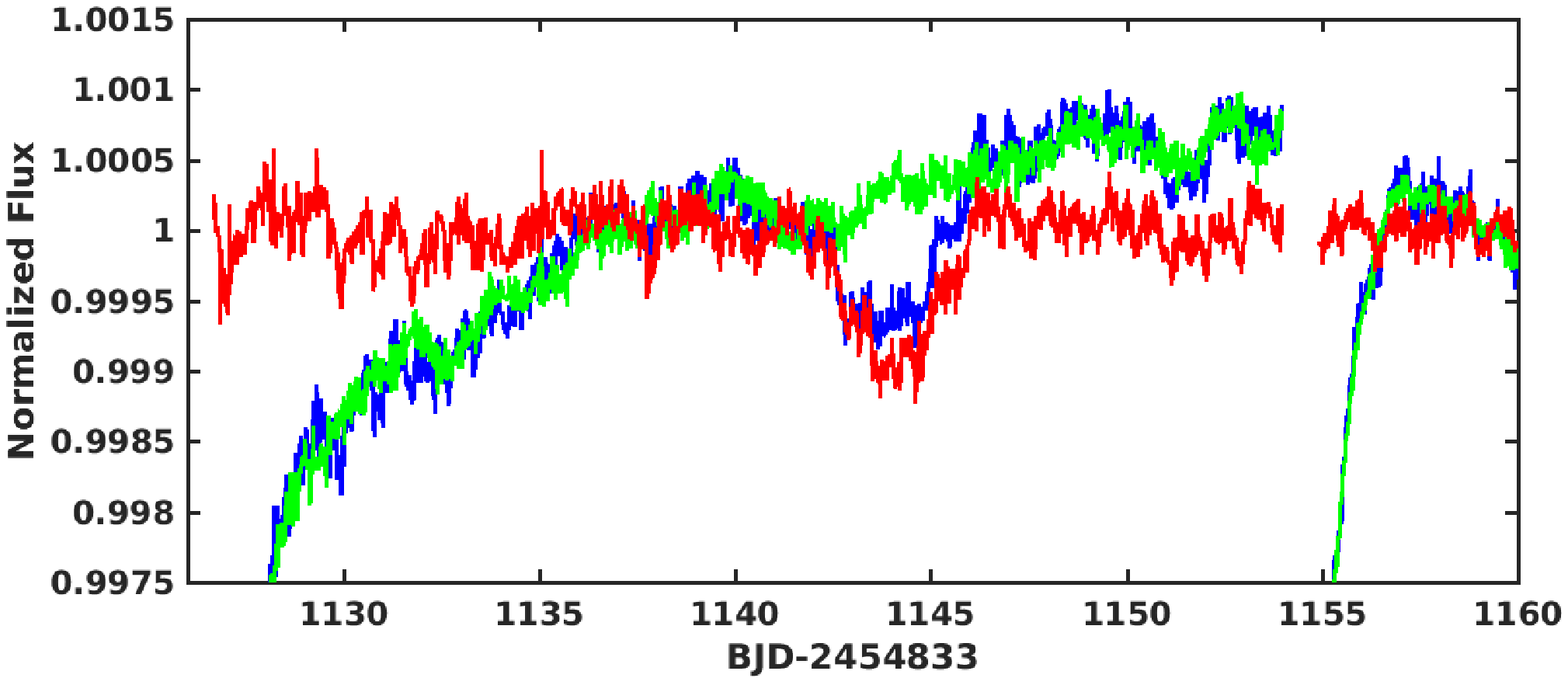}
\caption{\label{fig:CBV_ev12} Light curves at the time of the photometric events~A (upper panel) and B (lower panel).  
The curves show the detrended (red) and undetrended (blue) data sets. The fitted CBVs continuum is plotted with a green line.}
\end{figure}

Among the detected features, two events show a remarkable similarity in shape, duration and depth : 
the events~\#2 and \#13 in Table~\ref{tab:events}. Hereafter, we label these events as ``event~A'' and ``event~B''.  
The light curves of these two events are plotted in Fig~\ref{fig:CBV_ev12}. 
In this figure, we superimposed the raw SAPs, fitted CBVs and corrected SAPs, showing 
that the two photometric dips are real and not produced by the data analysis procedure. 

As indicated in Table~\ref{tab:events}, events~A and B were already noticed by \cite{Makarov2016} but they were suspected to be due 
to either PSF centroid modulation (event~A) or instrumental jitter (event~B). 
In section 4, we show that these events are of astrophysical origin and not related to instrumental systematics. 

\begin{figure}\centering
\includegraphics[height=80mm, angle=90]{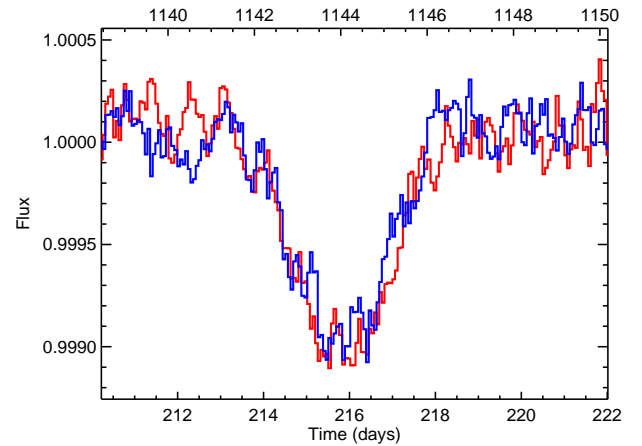}
\caption{\label{fig:events_superimposed} The 2~events superimposed with 3-pixels binning. The bottom x-axis shows the time at the first photometric event~A (red line). The top x-axis shows the time at the second photometric event~B (blue line), with a shift of 928.25~days relative to the bottom axis.}
\end{figure}

Fitting events A and B together, we derived a time separation between them of $\Delta t$=928.25$\pm$0.25 days. The errorbars on the flux were scaled to
obtain a reduced $\chi^2$ of 1. Shifting the second event light curve by -$\Delta t$, \textit{i.e.} on top of the first event light curve, 
we obtained the strikingly almost perfect superimposition of the two events as displayed in Fig.~\ref{fig:events_superimposed}.  

\begin{figure}\centering
\includegraphics[height=80mm, angle=90]{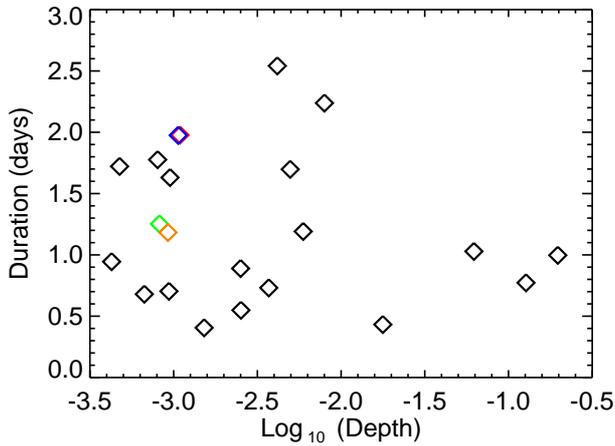}
\caption{\label{fig:depth_duration} Plot of the depth and duration of the 22~events cataloged in Table~\ref{tab:events}.
The events~A and~B are shown by red and blue symbols, respectively. 
The events~\#6 and~\#9 are shown by green and orange symbols, respectively. 
}
\end{figure}

\begin{figure}\centering
\includegraphics[height=80mm, angle=90]{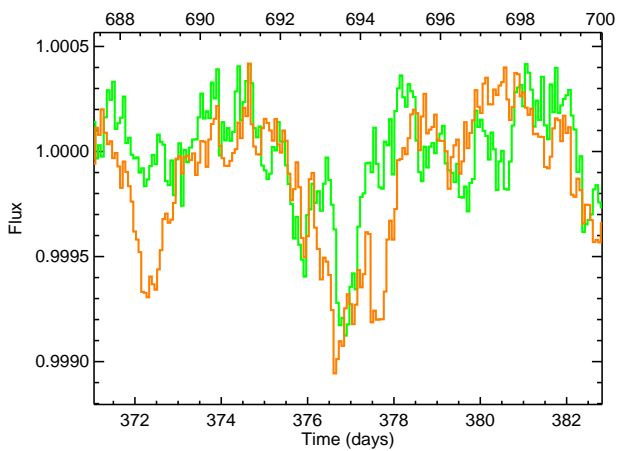}
\caption{\label{fig:events_6_9_superimposed} The events \#6 and \#9 superimposed with 3-pixels binning. The bottom x-axis shows the time at the first photometric event~\#6 (green line). The top x-axis shows the time at the second photometric event~\#9 (orange line), with a shift of 316.3~days relative to the bottom axis.}
\end{figure}

To characterize the similarity of these two photometric events and compare them among the 22~detected events, 
we plotted the depth and the duration of each of them (Fig.~\ref{fig:depth_duration}). 
The depths are measured between the continuum level fixed to 1 and the bottom of the light curve defined as the 5$^\text{th}$ lowest pixel. 
The durations are measured by calculating the second moments of the variations, 
which are then multiplied by $2\sqrt{2\ln 2}$ to roughly correspond to the full width at half maximum. 
While other photometric events show a wide diversity in duration and depth, 
events~A and~B are remarkably identical.
To emphasize this result, we superimposed the light curves of the photometric events~\#6 and \#9, 
which are, after the events~A and~B, the closest in the depth-duration diagram (Fig.~\ref{fig:events_6_9_superimposed}). 
It is clear that these photometric events do not show similar shapes of the light curves, as the events~A and~B do. 

The shape of the light curves of the events~A and~B can be obtained by fitting a simple 4-vertices polygon to each curve. 
We measure the quantities such as ingress, egress and centroid timings, slopes of the left and right wings, 
and transit depth (Table~\ref{tab:event_details}). These simple fits quantitatively confirm that the two events are strikingly similar.
The average duration of the two events from ingress to egress is measured to be 4.44$\pm$0.11 days. 
The bottom of the light curves are flat with a duration of about 1 day. The two slopes on each side of the flat bottom are straight, 
with comparable duration between 1.5 and 2~days. The right wings are steeper than the left wings, 
with respective slopes of about 700 ppm/day and -500 ppm/day. 
The transit depths are similar in both events at about 1010$\pm$40\,ppm. 

\begin{table}
\caption{\label{tab:event_details} The 4-vertices polygon parameters of the fit to the events~A and B light curves. 
We used a 4-degree polynomial to fit out the baseline, and assumed that the bottom of the lightcurve is flat. 
Beginning-of-ingress, centroid and end-of-egress timings are given in days past Kepler initial epoch at MJD 2454833. }
\begin{tabular}{@{}p{1.6cm}l@{~~}c@{~~~}c@{}}
                                            &                               & Event A                     & Event B \\
\hline

&&\\

 Continuum                                  &   $F_{\rm top}-1$ (ppm)       & 0.9$\pm$6.9                 &   1.3$\pm$6.5      \\ \\
\ldelim\{{4}{10pt}[Left wing\phantom{a}~~]  & slope ($10^{-4}$\,day$^{-1}$) & -4.94$\pm$0.24              & -5.24$\pm$0.30     \\
                                            & $\Delta t$ (day)              &  2.10$\pm$0.09              &  1.87$\pm$0.10     \\                
                                            & noise  (ppm)                  &  123                        &  148               \\
                                            & $t_{\rm ingress}$ (day)       &  213.32$\pm$0.06            &  1141.83$\pm$0.08  \\ \\
\ldelim\{{4}{10pt}[Bottom\phantom{aaa}~~]   & depth (ppm)                   & 1039$\pm$25                 &   978$\pm$16       \\ 
                                            & $\Delta t$ (day)              &  1.00$\pm$0.07              &  0.92$\pm$0.07     \\                
                                            & noise  (ppm)                  &  123                        &  109               \\
                                            & $t_{\rm centroid}$  (day)    &  215.93$\pm$0.05            &  1144.15$\pm$0.05  \\ \\
\ldelim\{{4}{10pt}[Right wing~~]            & slope ($10^{-4}$\,day$^{-1}$) &  7.41$\pm$0.28              &  6.18$\pm$0.20     \\
                                            & $\Delta t$ (day)              &  1.40$\pm$0.04              &  1.58$\pm$0.05     \\                
                                            & noise  (ppm)                  &    100                      &  142               \\
                                            & $t_{\rm egress}$   (day)     &  217.83$\pm$0.01            &  1146.19$\pm$0.03  \\ 
\hline
\end{tabular}
\end{table}

If real, these similar events could be the repeated observation of a same, identical, and periodic phenomenon. 
After checking for potential reduction artefacts and other systematics in the next section, 
we will discuss interpretations of this repeating event
in Sect.~\ref{sec:models}.  

\section{Possible bias}
\label{sec:bias}

\cite{Makarov2016} argued that some of the small amplitude features could be of instrumental or background origin. We thus carefully inspected the pixel tables collected by Kepler around the star's point spread function (PSF) and used to produce the raw SAP lightcurve about the epochs of the two events, in order to exclude any instrumental origin for these events, or the possibility of close background stars contamination. 

\subsection{Background stars}
\label{sec:backgnd}

The closest known stars in the field are Gaia\,2081900738645631744 (at 5.4" with $m_G$=18.1), referred to as Gaia-208 in the following, and the infrared sources 2MASS\,J20061551+4427330 (at 8.9" with $m_J$=16.1, $m_G$=18.9) and 2MASS\,J20061594+4427365 (at 11.83" with $m_J$=16.4, $m_G$=18.7). Their high visual magnitude measured by Gaia \citep{Vanleeuwen2017} implies a $\Delta V$$>$$6.4$ with KIC\,8462852 ($m_G$=11.7), \textit{i.e.} a flux ratio $<$0.3\%. 

As can be seen on Fig.~\ref{fig:pixel_map}, the pixels corresponding to the theoretical location of the 2 IR sources on the CCD channel are out of the aperture used to calculate the raw lightcurve. We see that KIC\,8462852's flux is smeared on an area about 10$\times$10\,arcsec$^2$ wide. Since the PSF of the two red stars is likely of similar extension, a bit less than half the flux of 2MASS\,J20061551+4427330 enters the PSF, while there is almost none for 2MASS\,J20061594+4427365. Consequently, any flux variation of these stars of order 100\% will contaminate the flux to a level lower than 0.1\%.  

However, since the PSF of Gaia-208 almost fully overlap with the PSF of KIC\,8462852, the photometric variations of this polluting star might induce variations in the lightcurve, but in any case not higher than 0.3\%.

At this stage, while we are able to exclude contamination from the two IR background stars, contamination from Gaia\,208 cannot be ruled out, although likely negligible. In section~\ref{sec:PSFmotion}, we show that no significant and correlated PSF motion of KIC\,8462852 is observed during the two events; this advocates for rejecting contamination from any background stars.

\subsection{Lightcurve of closest neighbour KIC\,8462934}

KIC\,8462934 is the closest bright star (about 89" with V$\sim $11.5) to KIC\,8462852 (V=12) with a recorded lightcurve in the Kepler Database. Applying our previously introduced detrending method, we recovered a detrended lightcurve using the 13 first CBVs of each quarter/channel, as was applied to KIC\,8462852's lightcurve (see Section~\ref{sec:kepler_data}). We found no peculiar behaviour, neither strong nor shallow absorptions similar to what observed on KIC\,8462852. No features were detected in the lightcurve beyond a few 10$^{-4}$ in normalized flux. This indicates that the CBVs derived by the Kepler pipeline did not miss any small local variations in \textit{e.g.} pixel sensitivity in the CCD-channel. Therefore, events A and B are indeed features from the local pixels in the photometric aperture shown in Fig.~\ref{fig:pixel_map}.

\subsection{KIC\,8462852's PSF motion on CCD pixels}
\label{sec:PSFmotion}

\cite{Makarov2016} studied the correlation of the PSF centroid motion and the flux dimming in KIC\,8462852. They show that a few features could be artifact of background objects occultation, or instrumental jitter. 

\begin{figure}\centering
\includegraphics[width=80mm, clip=true, trim=0 0 0 0]{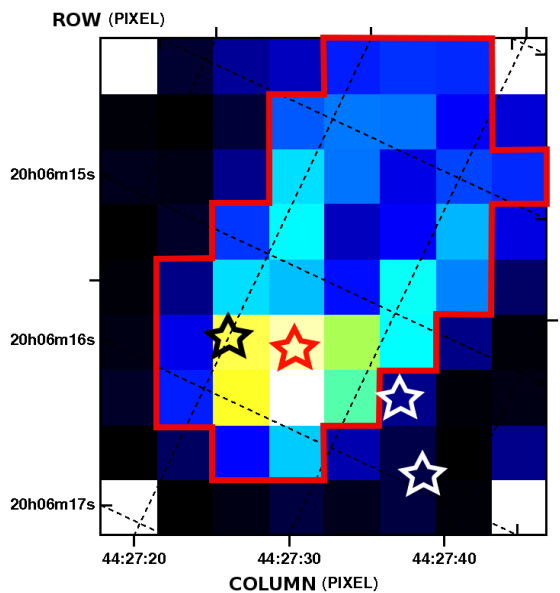}
\caption{\label{fig:pixel_map} Pixel table of flux around KIC\,8462852, collected during quarter \# 2 on CCD-channel 32. The colors are logarithmically scaled with the flux. The aperture is displayed in solid red line. A right ascension and declination grid is superimposed in dotted black lines. The theoretical position of KIC\,8462852 on the CCD is depicted as a red star; the position of the two faint IR sources 2MASS J20061551+4427330 \& 2MASS J20061594+4427365 are marked by white stars; and Gaia\,2081900738645631744 appears as a black star.}
\end{figure}

\begin{figure}\centering
\includegraphics[width=100mm, clip=true, trim=0 0 0 0]{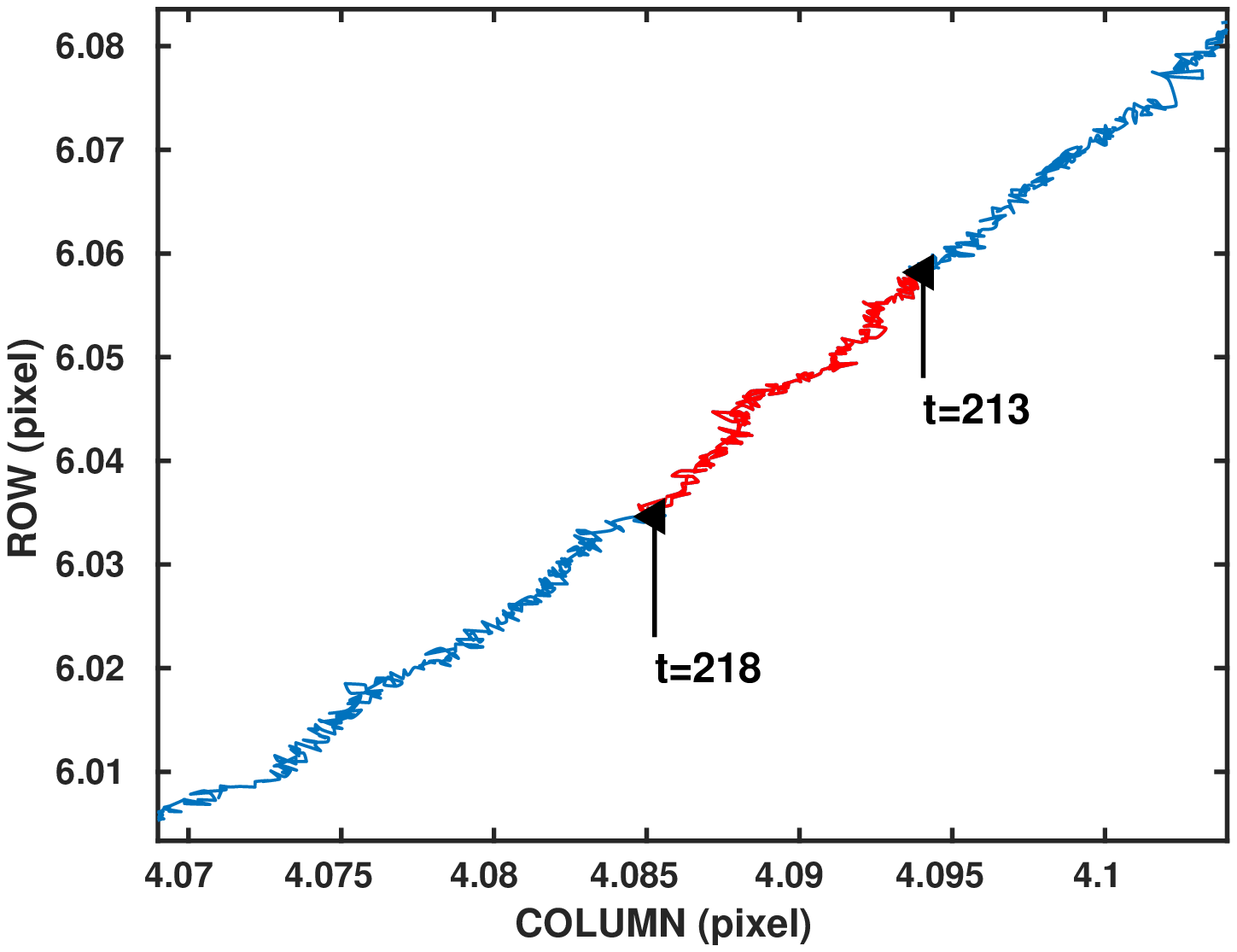}
\includegraphics[width=100mm, clip=true, trim=0 0 0 0]{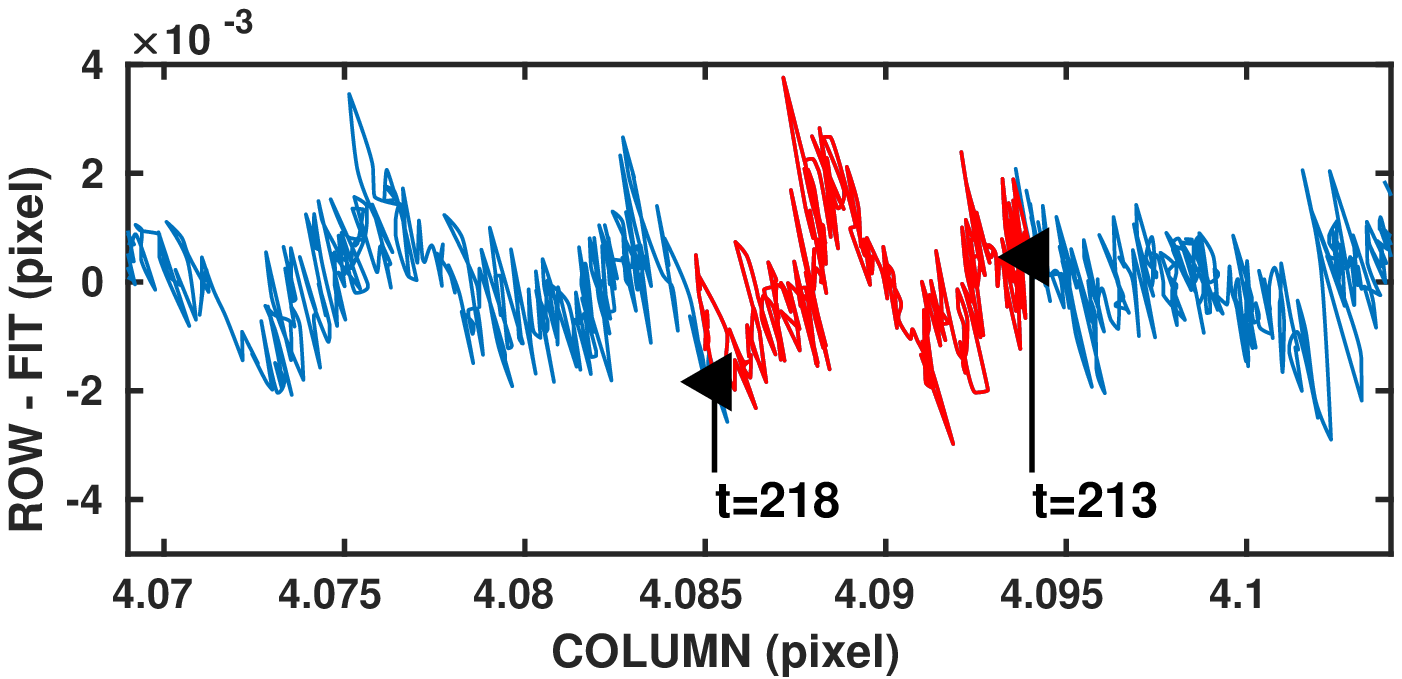}
\caption{\label{fig:PSF_motion} \textbf{Upper panel:} position of the PSF centroid around event A, from epochs 208 to 225 (in blue), and highlighted in red, from ingress (213) to egress (220). Time arrow goes from top right to bottom left. \textbf{Lower panel:} PSF centroid motion around the 4th degree polynomial fit of the main trend in the upper panel. The modulation period is about 3 days.}
\end{figure}

Repeating a similar analyzis on the pixel images collected by Kepler at each cadence, such as presented in Fig.~\ref{fig:pixel_map}, we derived about the epochs of each identified event the centroid motion of KIC\,8462852's PSF. Apart from an expected slow shift (0.004 pixel\,day$^{-1}$) of the star's location on the CCD channel during the quarter, we observed that the centroid oscillates with a period of about 3 days and an amplitude of at most a few 0.001 pixel (Fig.~\ref{fig:PSF_motion}). This oscillation is likely related to a vibration mode of the instrument. Consistently, it was found that a few pixels show shallow flux modulations of a few 0.1\% in correlation or anti-correlation with the PSF centroid oscillations. 

Impressively, these instrumental modulations exactly cancels out. We found no counterpart for these modulations in the raw lightcurve, demonstrating the excellent quality of the flat field determination made by Kepler on the aperture. Since the amplitude of the 3 days-modulation is similar to the amplitude of the two events discussed in this paper, instrumental PSF variations cannot be at their origin. Indeed, in such case, we would have observed 0.1\% deep 3 days-modulation rather than only single dips. 

We have seen in Section~\ref{sec:backgnd} that the closest background star, Gaia-208, is located at 5.4" from KIC\,8462852. This is a bit more than 1 pixel apart, on the aperture (Fig.~\ref{fig:pixel_map}). The luminosity variation due to the occultation of a third of Gaia-208 stellar disk would be close to 0.1\%, leading to a PSF centroid variation of about 10$^{-3}$ pixels. We can exclude from Fig.~\ref{fig:PSF_motion} a PSF centroid variation of this amplitude during event A, between day-213.3 and day-217.8. 

Repeating this analysis for event B led to the same conclusion. We thus exclude for both events any significant PSF motion correlated with the lightcurve, eliminating background star pollution, and instrumental variations as possible origin.

\subsection{Local pixel variations}
\label{sec:final_check}

As a final check, we verified the collective variations of the local flux in each pixel around the PSF of KIC\,8462852 at different times between the beginning and the end of both events. 
It clearly appeared that the whole image of the star was fainting during the dimming events, thus confirming that its origin is neither related to background objects occultation, nor associated to any instrumental PSF motion.


\section{Models}
\label{sec:models}

We can try to explain the repeated photometric event, observed 928~days apart. 
The observed variations correspond to a dip in the star brightness by about 0.1\%, which lasted for about 4.4~days. 
A decrease in the star brightness (moreover in a context of a star showing multiple photometric variations, 
always in the form of brightness decrease) suggests an explanation by the transit of a partially occulting body. 
With that in mind, the duration of the event ($\sim$5~days) is puzzling. 
With a possible period of 928~days, and assuming a mass of 1.4~solar mass for the F3V central star, 
the corresponding semi-major axis is 2.1\,au and the orbital velocity on a circular orbit is 24.4\,km\,s$^{-1}$.
 At this transiting velocity, the maximum transit time in front of a $R_*=1.3R_\odot$ star is about 10.3~hours.
Even on an highly eccentric orbit and observed at apoastron, the transit of a body on a 928-days period orbit cannot last longer than 
14.6~hours.
Therefore, the photometric events of 4.4-days can be explained by the transit of an occulting body 
only if this body is significantly larger in size that the star ; in this case, the duration of the transit is 
related to the size of the object itself.  

The main scenario for explaining the other deeper dips in the KIC\,8462852 light curve invokes the transit 
of trains of extrasolar comets \citep{Boyajian2016, Bodman2016} or planet fragments \citep{Metzger2016}. 
In fact, the photometric variations observed in KIC\,8462852 light curve look like the spectroscopic variations 
observed in $\beta$\,Pictoris, which can last several days and are interpreted 
by the transit of exocomets \citep{Ferlet1987, Lagrange1992, Vidal-Madjar1994, Kiefer2014a}. 
We will explore this scenario in Sect.~\ref{sec:comets string}.

Nevertheless, keeping the idea of a transiting body, we can imagine another possible scenario 
to explain the repeated photometric events~A and~B. The straight ingress and egress slopes, 
and the flat bottom  of the light curve point toward the possibility that the transiting body 
can be a single body with a simple shape. 
Acknowledging that the Hill-spheres of a massive planet can extend to several stellar radius in size, 
the transit of a ring system surrounding a giant planet could explain the observed photometric event~A and~B,  
as done for the light curve of 1SWASP J140747.93-394542.6 (also named J1407), 
an old star in the Sco-Cen OB association \citep{Kenworthy2015}; see also \cite{Lecavelier2017} and \cite{Aizawa2017} for other case studies of exo-planetary ring systems. 
This scenario will be discussed in Sect.~\ref{sec:ring model}.

\subsection{The comets string model}
\label{sec:comets string}

\begin{figure}\centering
\includegraphics[width=80mm]{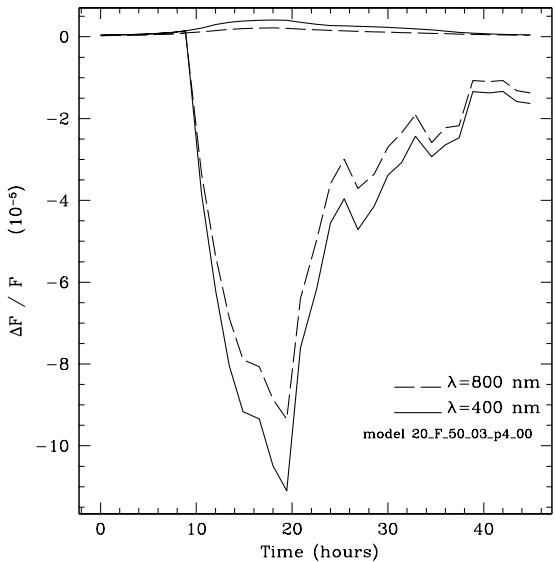}
\caption{\label{fig:model_1999} The modeled exocomet transit light curve from \cite{Lecavelier1999b} 
for the case '20\_F\_50\_03\_p4\_00', corresponding to an exocomet orbiting an F star with a periastron of 0.3\,au, 
a longitude of periastron of 90\degr, and a production rate of 10$^5$~kg/s at 1\,au.}
\end{figure}

In the exocomets scenario, the duration of the transit event in the light curve implies that 
several comets passed in front of the star, within an extended string long of several millions of kilometers. 
While we do not aim at exploring the whole range of possibilities to fit the events~A and~B light curves, 
we could use some of the exocomet tail transit signatures given \textit{e.g.} in the library of \cite{Lecavelier1999b}
to show that a generic transit model of a few trailing exocomets can easily provide a satisfactory fit to the data.  

As a reference light curve, we decided to use the light curve labeled '20\_F\_50\_03\_p4\_00' in \cite{Lecavelier1999b}, and plotted on Fig.~\ref{fig:model_1999}. 
It is obtained through the simulation of cometary tails orbiting an F star with a periastron of 0.3\,au, a longitude of periastron of 90\degr, and a production rate of 10$^5$~kg/s at 1\,au~\citep{Lecavelier1999a}. It assumes a grain size distribution given by 
$dn(s) = (1-s_0/s)^m s^{-n} \, ds$, with $s_0$=0.05\,$\mu$m, $n$=4.2, $m$=$n(s_p-s_0)/s_0$, and peaking at $s_p$=0.2\,$\mu$m. This distribution is derived from observations  in solar system comets at less than 0.5~\,au from the Sun. The physical model used to calculate the photometric transit signatures of exocomet tails is discussed in depth in \cite{Lecavelier1999a}.

The choice of the characteristics (the orbit and the dust production rate) of the specific exocomet 
for the reference light curve is not critical, because all the transit light curves show a similar triangular shape. 
At a fixed distance to the star, the transit depth of an individual light curve is constrained by the dust production rate, and the duration is mainly related 
to the longitude of the periastron. The depth of the global light curve resulting from the transit 
of a string of several exocomets therefore depends on the production rate of each exocomet. However, 
the duration of the global light curve is not constrained by the duration of each individual transit, but by the spread of the transit time of each exocomets.

To simplify the fit to the data, we approximated the reference light curve of a single comet 
by a piecewise linear function. Each individual exocomet lightcurve is defined by two parameters: 
the time of mid-transit $T_0$, and the maximum occultation depth, $\Delta F/F$. 
Exploring the library of \cite{Lecavelier1999b}, we find that 
in the range 10$^4$-10$^6$\,kg/s the maximum occultation depth is related to the dust production rate, $\dot M$, by 
$\log_{10} \dot M/(1\,{\rm kg\,s^{-1}}) =5+1.25\times\log_{10}(\Delta F/F /10^{-4})$. 

We fitted the average light curve of events~A and~B 
with a combination of several individual light curves defined by $T_{0k}$
and $\dot M_k$ for each comet $k$ of the string. With N comets in the string, the total number of parameters reaches $2N+1$ with 2 parameters per 
comet and one for the baseline level (slightly larger than 1). 
Given the possibly large number of parameters, we used a Markov-chain Monte Carlo (MCMC) algorithm as a fitting procedure.

\begin{figure}\centering
\includegraphics[height=80mm, clip=true, angle=90]{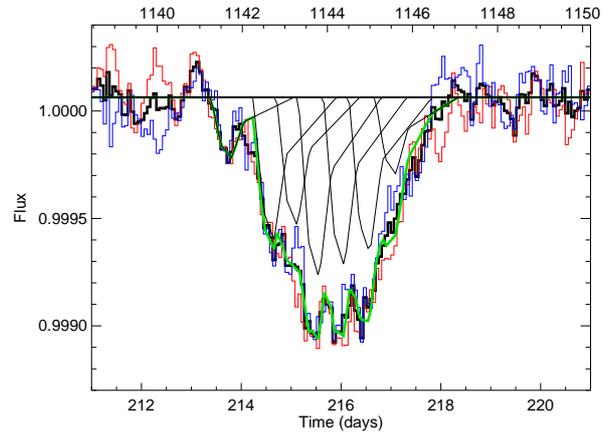}
\caption{\label{fig:frag_comet_model_7}
Fit to the light curve using a string of 7~exocomets.
The light curve of each exocomet is given by the thin black lines.
The data of the events~A and~B are plotted with the red and blue thin lines,
and the co-addition of the two light curves is given by the thick black line.
The best fit is plotted with the thick green line.}
\end{figure}

The best fit is obtained for 7 comets, including the feature at the top of the signature left wing. 
It is plotted in Fig.~\ref{fig:frag_comet_model_7} with the parameters
given in Table~\ref{tab:comet_model_57} and plotted in Fig.~\ref{fig:frag_comet_prod}.
The dust production rates obtained for the comets are typical of Hale-Bopp type comets
in the Solar System, \textit{i.e.} between 10$^5$ and 10$^6$\,kg\,s$^{-1}$ \citep{Huang2000}.

\begin{figure}\centering
\includegraphics[height=80mm, clip=true, angle=90]{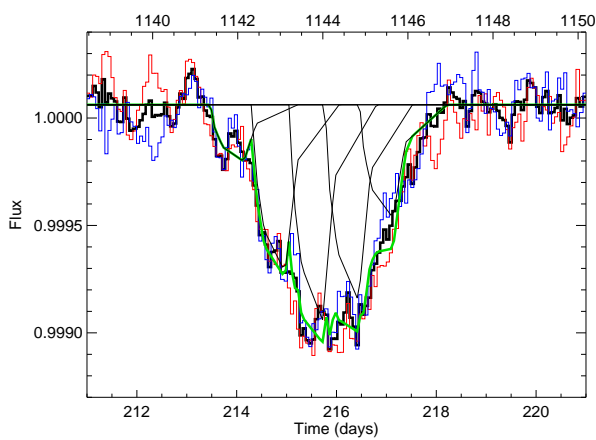}
\caption{\label{fig:frag_comet_model_5}
Same as Fig.~\ref{fig:frag_comet_model_7} using a string of 5 exocomets. }
\end{figure}

\begin{figure}\centering
\includegraphics[height=80mm, angle=90, clip=true,trim=0 0 0 0]{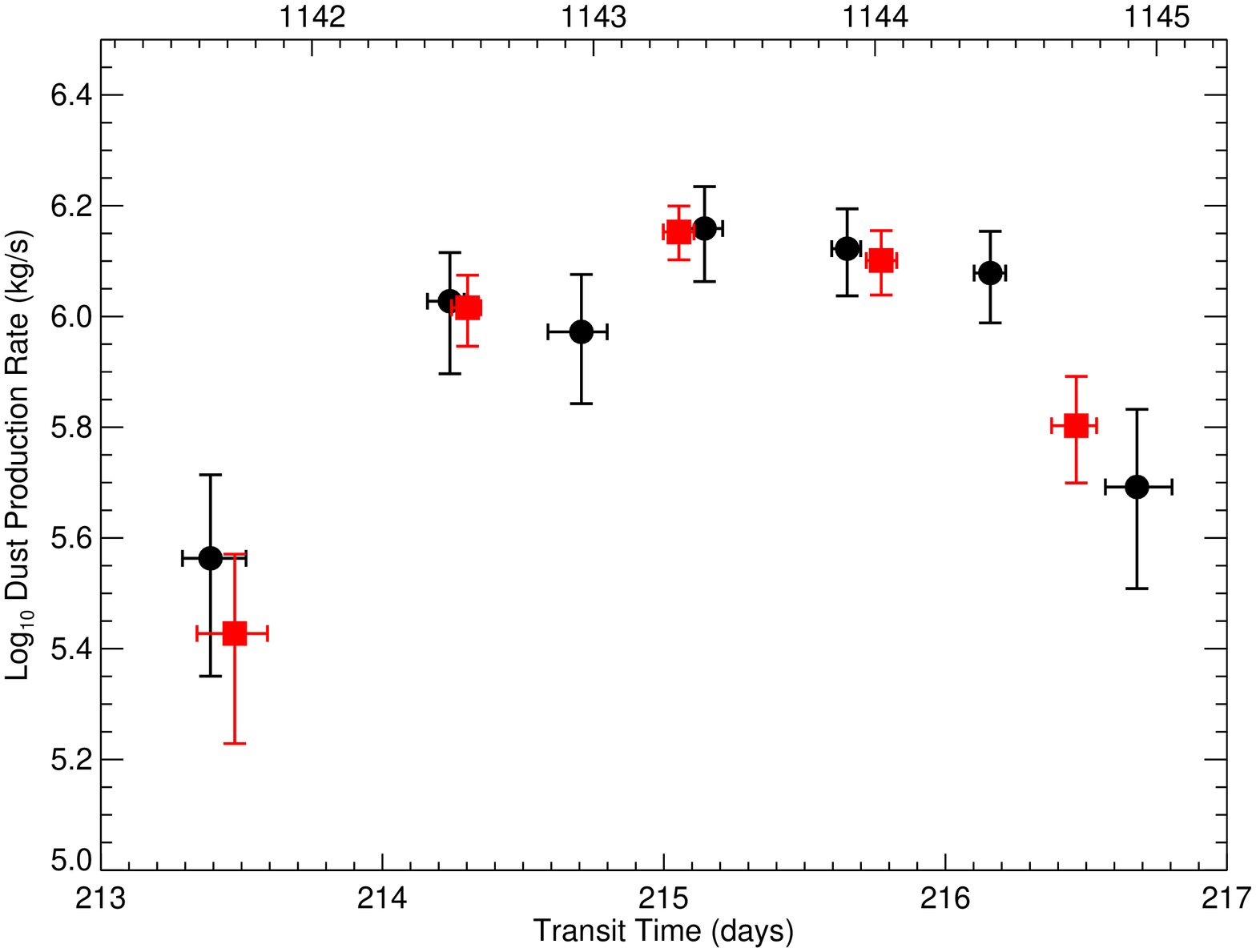}
\includegraphics[width=90mm, angle=0, clip=true,trim=0 0 0 0]{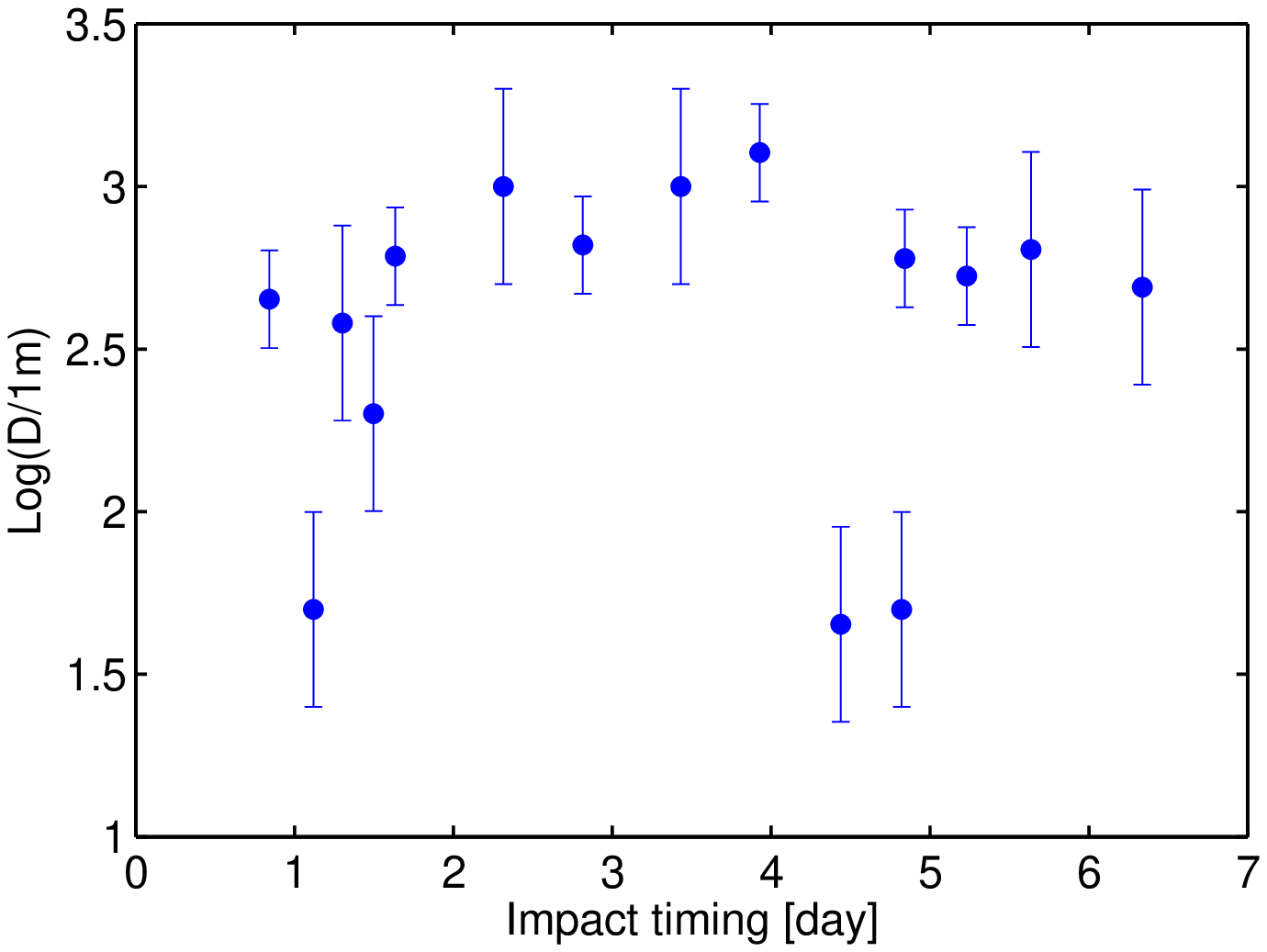}
\caption{\label{fig:frag_comet_prod} \textbf{Top:}
The dust production rates and the transit times of the transiting bodies for the 7~comets model
(black dots; Figure~\ref{fig:frag_comet_model_7})
and the 5~comets model (red squares; Figure~\ref{fig:frag_comet_model_5}).
\textbf{Bottom:} Shoemaker-Levy 9 fragments diameter versus epochs of impact with Jupiter's atmosphere.}
\end{figure}

If we consider we are actually overfitting stellar variations, 
we could accept a poorer fit with residuals in the order of the mean amplitude of the stellar variations. 
In this case, 5~comets are enough to fit satisfyingly the average light curve.
An example of such a fit is shown in Fig.~\ref{fig:frag_comet_model_5} with the parameters
given in Table~\ref{tab:comet_model_57} and plotted in Fig.~\ref{fig:frag_comet_prod}.
Here we used a different longitude of periastron of 112.5\degr\ and
a different grain size distribution labeled '50' in \cite{Lecavelier1999b},
peaking at 0.5\,$\mu$m.
This shows that the observations cannot constrain the comets properties
and that the comets model can easily explain the data without any fine tuning of parameters.
Therefore, the values given in Table~\ref{tab:comet_model_57} should not be considered
as measurements on existing bodies, but as possible values for a generic model of a string of exocomets.

Interestingly, both resulting models are reminiscent of the case of the Solar System comet Shoemaker-Levy 9 (SL9). 
Figure~\ref{fig:frag_comet_prod}, bottom-panel, shows the distribution of diameters, $D$ (in log-space), with respect to timing of impact with Jupiter of all 21~fragments of SL9 \citep{Hammel1995, Chodas1996, Crawford1997}. Since the dust production rate is proportional to the surface of the nucleus (all other things equal), $\log D$ of SL9 fragments could be compared to $\log \dot M$ of events A and B comets (Fig.~\ref{fig:frag_comet_prod}, top-panel). We see that in both cases, the distribution of size (evaporation rate) is mainly flat with decreasing size of the comet nuclei at the head and tail of the fragments string. This tentatively suggests events A and B could be the break-up remnants of a bigger body along its orbit. If the periodicity of this transit is later confirmed, non-gravitational effects should be properly modelized to take into account a slow relative drift of the fragments.


\begin{table}\centering
\caption{\label{tab:comet_model_57} Best fit parameters for the models with the transit of 7 or 5~exocomets. The error bars correspond to 3-$\sigma$ and have been estimated using MCMC. The central transit time, $T_0$, are given for the event~A, a constant of 928.25\,days must be added for the event~B. }
\begin{tabular}{lcc}
Comet  & Transit time & Dust production rate \\
       & $T_0$        & $\dot M$ \\
       & (day)        &   ($\log$ kg\,s$^{-1}$)     \\
\hline
\multicolumn{3}{c}{\it 7 comets model} \\
\\
  1&    213.39  $\pm$  0.11 &      5.56  $\pm$  0.18\\
  2&    214.24  $\pm$  0.07 &      6.03  $\pm$  0.11\\
  3&    214.71  $\pm$  0.11 &      5.97  $\pm$  0.12\\
  4&    215.14  $\pm$  0.06 &      6.16  $\pm$  0.09\\
  5&    215.65  $\pm$  0.05 &      6.12  $\pm$  0.08\\
  6&    216.16  $\pm$  0.06 &      6.08  $\pm$  0.08\\
  7&    216.68  $\pm$  0.12 &      5.69  $\pm$  0.16\\
\hline 
\multicolumn{3}{c}{\it 5 comets model} \\
\\
  1&    213.48  $\pm$   0.13 &      5.43  $\pm$  0.17\\
  2&    214.30  $\pm$   0.05 &      6.02  $\pm$   0.06\\
  3&    215.05  $\pm$   0.05 &      6.15   $\pm$  0.05\\
  4&    215.77  $\pm$   0.05&       6.10  $\pm$   0.06\\
  5&    216.46  $\pm$   0.08&       5.80  $\pm$   0.10\\
 \hline
\end{tabular}
\end{table}



\subsection{The planetary ring model}
\label{sec:ring model}

Here we discuss another possible scenario consisting in the transit of a giant ring system 
surrounding a planet with a 928-days orbital period (2.1\,au semi-major axis). 
Indeed, a ring system can be stable within half a Hill-sphere radius of a planet. 
Around a massive planet the Hill-sphere can extend up to several stellar radii in size ; therefore rings \citep{Kenworthy2015, Lecavelier2017, Aizawa2017} or dust envelope like \textit{e.g.} Fomalhaut\,b \citep{Kalas2008} can be large enough
that the transit duration can reach up to a few days. For instance, the Hill-sphere of Jupiter extends up to 0.34\,au (73\,$R_\odot$). 

\begin{figure}\centering
\includegraphics[width=80mm, clip=true]{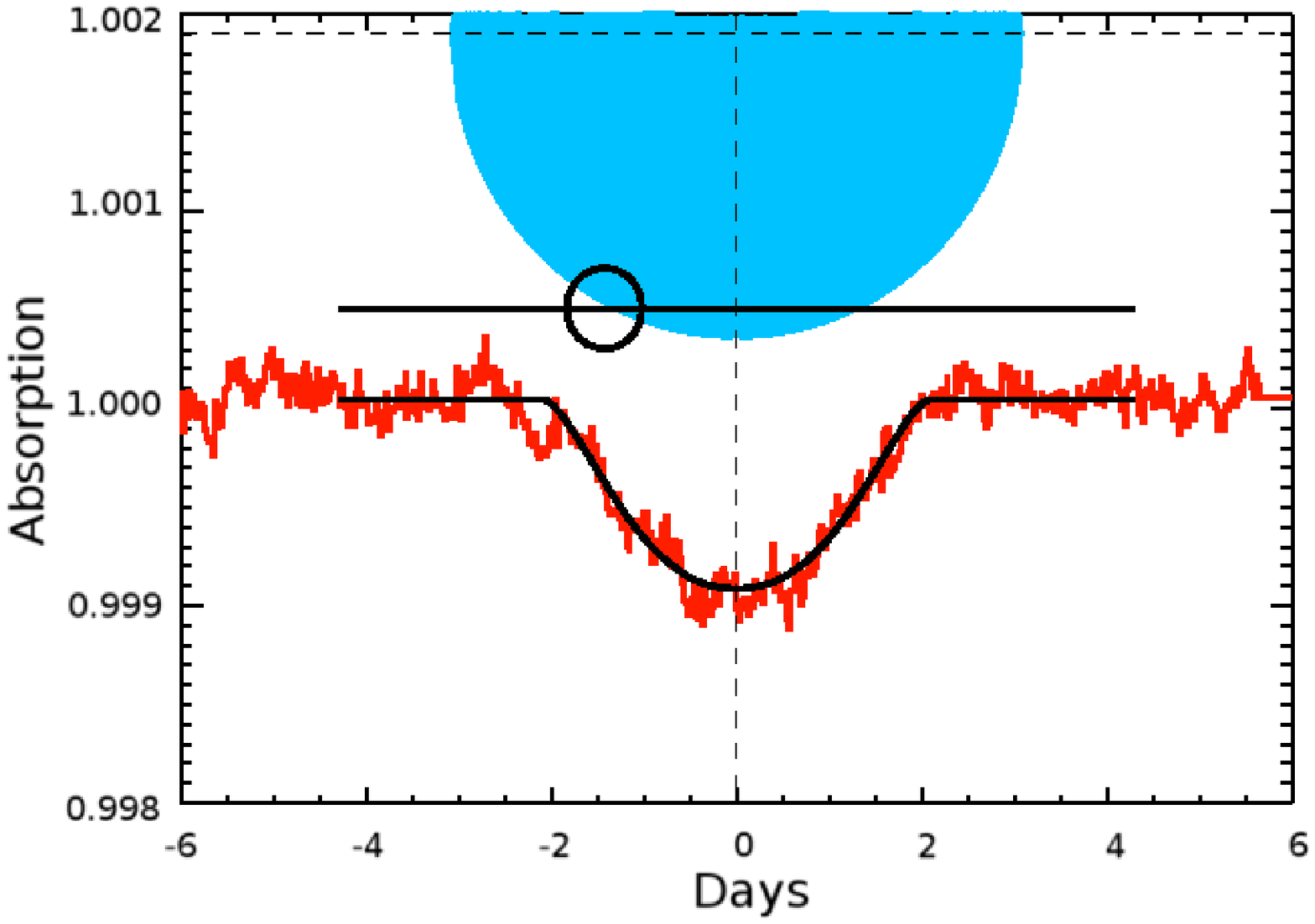}
\includegraphics[width=80mm, clip=true]{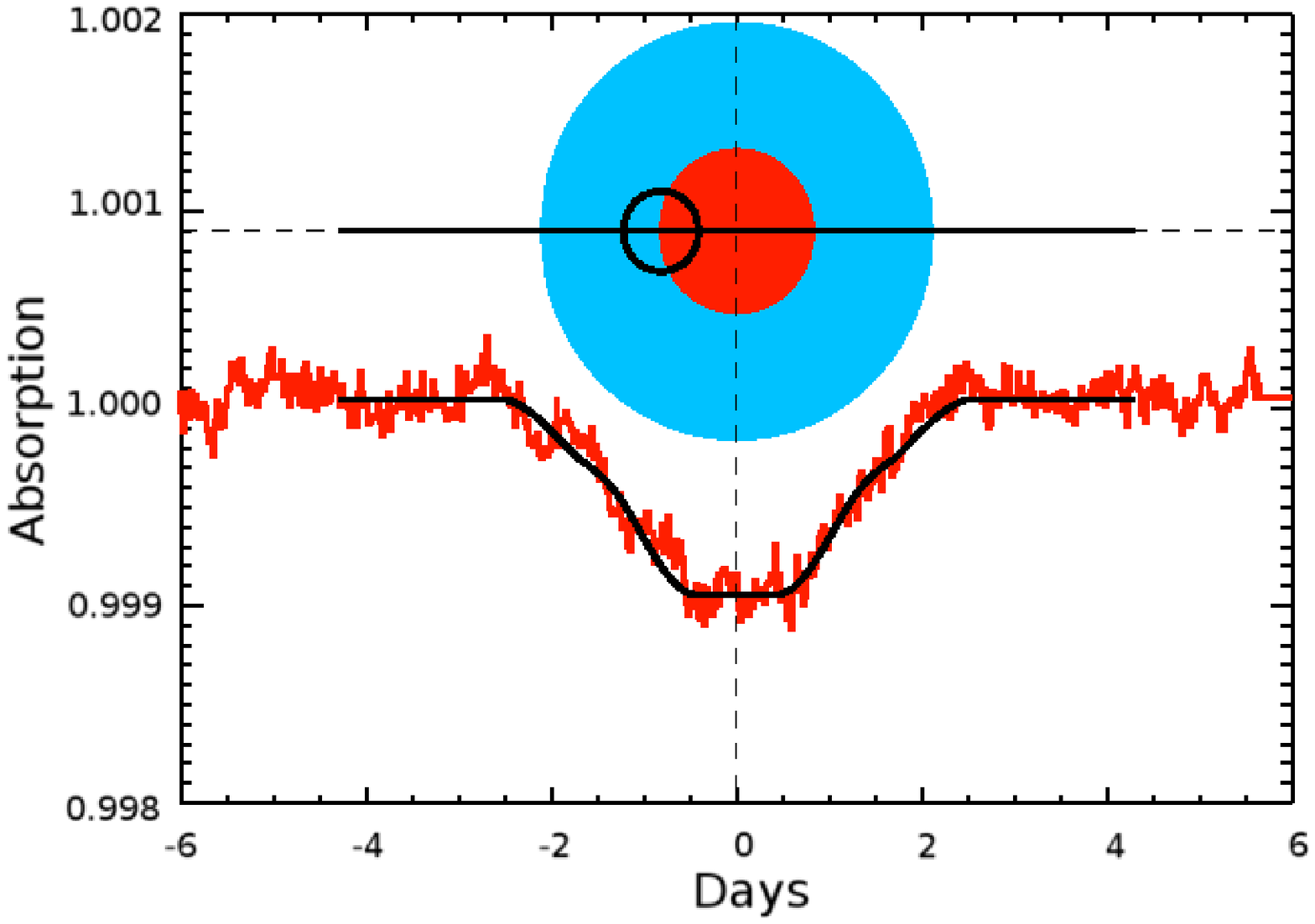}
\caption{\label{fig:ring_models} Fit to the average light curve of events~A and~B (in red) 
by the two models of KIC\,8462852's occultation by a planetary ring, 
as presented in Section~\ref{sec:models} (black curves). 
\textbf{Upper panel:} Homogeneous circular ring with a constant opacity and a non-zero impact parameter. 
\textbf{Lower panel:} Circular ring with a constant opacity in the center (red area) and a decreasing opacity with distance following a $r^{-\alpha}$ law in the external ring 
(blue area). The impact parameter is fixed to 0. 
The black circle figures the edge of stellar disk.} 
\end{figure}

To model this scenario, we take the reference frame linked to the planet, and consider that the star transits behind the rings. To simplify the problem, we assume that \textit{i)} the planet moves on a circular orbit at 2.1\,au (v$_\text{transit}$=24.4\,km\,s$^{-1}$); and \textit{ii)} that the rings are seen face-on. We consider two simple models of rings, and fitted them to the data using Levenberg-Marquardt minimization of the $\chi^2$:
\begin{itemize}
\item[1.] The first model consists in a large circular homogenous, constant opacity, ring with a non-zero impact parameter of the star's trajectory behind the ring during the transit (Fig.~\ref{fig:ring_models}, left panel). In this case, the signature of the transit is round-shaped. 
The data are best fitted with a ring exterior diameter of 8.8\,$R_\star$, an impact parameter of 8.5\,$R_\star$ and an extinction $\tau$=0.0014. Nonetheless, this model
do not provide a good fit to the data, which show straight wings and a flat bottom. \\
\item[2.] In the second model, the ring is made of an inner core of  
constant opacity for $r$$<$$R_{\rm const}$ and an external ring with an extinction decreasing with the distance to the star 
following $\propto$ \!$r^{-\alpha}$ for $r$$>$$R_{\rm const}$ 
(Fig.~\ref{fig:ring_models}, right panel). 
As can be seen in the figure, this model provides a much better fit to the data. 
Using a zero impact parameter, the best fit is found with an outer radius of 4.86$\pm$0.15\,$R_\star$, 
an interior core of radius $R_{\rm const}$=1.91$\pm$0.03\,$R_\star$ 
with constant extinction $\tau$=(9.9$\pm$0.1)$\times$10$^{-4}$, 
and an extinction parameter $\alpha$=1.70$\pm$0.06.
\end{itemize}

We tried more sophisticated models by introducing elliptical rings, non-zero impact parameter and a non-zero position angle of the ellipse major-axis with respect to the transit direction (model \#3 in Table~\ref{tab:ring_models}).  
The improvement of the fit is significant but only indicates that the rings as seen for Earth are likely elliptic ($e$>$0.8$) and not aligned with the transit direction. This is in accordance with the observed asymmetry on the slopes of the left and right wings, as explained in Section~\ref{sec:2events}. Since the projection of an inclined circle is an ellipse, the eccentric solution corresponds to a circular ring system inclined with respect to the plane-of-the-sky at angle $\theta$($=$$\arcsin e$)$>$$53^\circ$.

Interestingly, \cite{Ballesteros2017} recently proposed that the two main dips (D800 and D1500) of KIC\,8462852 could be related to a ring planet on a 12 years orbit, with trailing trojans at the L5 point. If true, KIC\,8462852 might be the first exoplanetary system with two ring planets detected. 

\begin{table}\centering
\caption{\label{tab:ring_models} Table of $\chi^2$ and BIC of the different ring models proposed in the text. $\tau$ is the extinction.}
\begin{tabular}{lcccc}
 Models               &  $N_{\rm param}$   &  $N_{\rm DOF}$   & $\chi^2$  & BIC\\
         \hline 
\#1. Circular ring    &      \multirow{2}{*}{5}              &     \multirow{2}{*}{430}         &    \multirow{2}{*}{590.0} &  \multirow{2}{*}{620.4}\\
\textit{$\tau$ constant} & \\ \\
\#2. Circular rings   &      \multirow{3}{*}{6}              &     \multirow{3}{*}{429}         &    \multirow{3}{*}{550.1}  &  \multirow{3}{*}{586.6}\\
\textit{$r>R_{\rm const}$, $\tau\propto r^{-\alpha}$}  \\ 
\textit{$r<R_{\rm const}$, $\tau$ constant} \\ \\
\#3. Elliptic rings  &       \multirow{7}{*}{9}              &     \multirow{7}{*}{426}         &    \multirow{7}{*}{515.6}   &   \multirow{7}{*}{570.3}  \\
\textit{$r>R_{\rm const}$, $\tau\propto r^{-\alpha}$}  \\ 
\textit{$r<R_{\rm const}$, $\tau$ constant} \\
$e>0.8$ \\
\textit{Impact parameter $\neq$ 0} \\
\textit{Position angle $\neq$ 0} \\
\hline       
\end{tabular}
\end{table}

\section{Observing the future events}

With the last event on BJD 2455977.15, and assuming periodicity with $P$=$t_{B}-t_A$=928.25$\pm $0.25\,days, the phenomenon is expected to repeat itself every $t_{B} + N \times P$. The occurence timing closest to the present date is for $N$=$2$ (event D) with
\begin{align}
T_{D} &= 2457833.65 \pm 0.80  \\ &\text{or between the 20th of March 2017 at 07:55 UT} \nonumber \\ & \quad\text{ and the 21st of March 2017 at 23:17 UT.} \nonumber 
\end{align}

The beginnning-of-ingress and end-of-egress timings were also estimated. Table~\ref{tab:timing_events} summarizes these informations.
\begin{table}\centering
\caption{\label{tab:timing_events} Timing and ephemeris of transit events with $P$=$928.25$\,days starting from event B at $t_B$=1144\,days past Kepler initial epoch at MJD 2454833.}
\begin{tabular}{lll}
                     &   \multicolumn{1}{c}{MJD} &   \multicolumn{1}{c}{UT date}        \\
\hline
\multicolumn{3}{c}{\textbf{Most recent event in the past at $t_B + 2 \times P$ (event D)}}  \\ \\
$T_\text{ingress}$   &  $2457832.40$$\pm $$0.70$ &  19/03/17 (04:48) \\
                     &                           &  $\rightarrow$ 20/03/17 (14:24) \\
$T_\text{centroid}$  &  $2457833.65$$\pm $$0.80$ &  20/03/17 (07:55) \\
                     &                           &  $\rightarrow$ 21/03/17 (23:17)  \\
$T_\text{egress}$    &  $2457835.70$$\pm $$0.60$ &  22/03/17 (14:24) \\
                     &                           &  $\rightarrow$ 23/03/17 (19:12)  \\ \\
\multicolumn{3}{c}{\textbf{Next event in the future at $t_B + 3 \times P$ (event E)}}  \\ \\
$T_\text{ingress}$   &  $2458760.65$$\pm $$0.74$ &  03/10/19 (09:30) \\
                     &                           &  $\rightarrow$ 04/10/19 (21:30) \\
$T_\text{centroid}$  &  $2458761.90$$\pm $$0.84$ &  04/10/19 (13:00) \\
                     &                           &  $\rightarrow$ 06/10/19 (06:00) \\
$T_\text{egress}$    &  $2458763.95$$\pm $$0.64$ &  06/10/19 (19:00) \\
                     &                           &  $\rightarrow$ 08/10/19 (02:30) \\
\hline
\end{tabular}
\end{table}

We planned observing KIC\,8462852 between the 19th of March and the 23rd of March 2017 in photometry and/or spectroscopy. Unfortunately, 
HST and Spitzer were both unable to point at KIC\,8462852 on these dates. State-of-the-art ground-based photometry 
is not sensitive and stable enough to confirm a 0.1\% deep transit signature lasting several days. Ground-based spectroscopy has been tried, since in case of exocomet transit, variable Na\,I or Ca\,II features could be expected in the KIC\,8462852 spectrum \citep{Kiefer2014a, Kiefer2014b, Beust1990, Ferlet1987}. We therefore planned observations of the star with the SOPHIE spectrograph installed on the 1.93m telescope of Observatoire de Hautes-Provence~\citep{Perruchot2008,Bouchy2009} between the 15th and the 26th of March 2017.

\begin{figure}\centering
\includegraphics[width=90mm, clip=true]{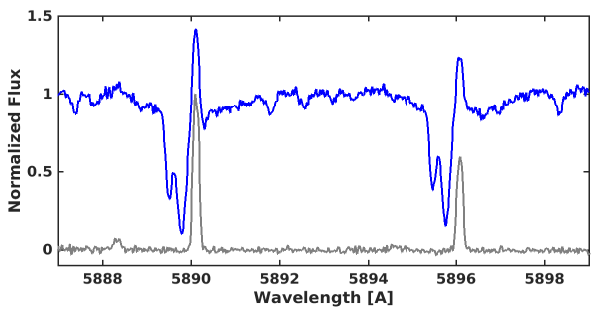}
\caption{\label{fig:NaI_spectrum} Na\,I spectra of KIC\,8462852. In blue, the average of the spectra collected through Fiber A of the SOPHIE spectrograph, on the 15, 16, 17 and 19th of March 2017. In grey, the average sky-background spectrum taken simultaneously with each star's spectrum on Fiber B. The emission line seen on Fiber A and B is clearly identified as geocoronal sodium emission. The double peak feature on the left of the telluric emission is most probably of interstellar absorption origin, since no counterpart is observed in the Ca\,II spectrum at the star radial velocity (see Fig.~\ref{fig:CaII_spectrum_compared}). }
\end{figure}

\begin{figure}\centering
\includegraphics[width=90mm, clip=true]{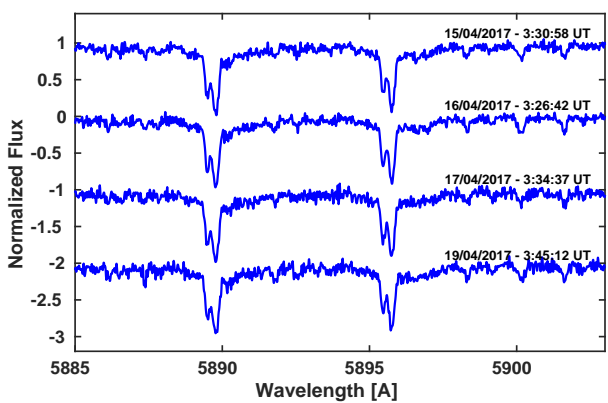}
\caption{\label{fig:NaI_spectrum_compared} Comparison of the Na\,I spectra of KIC\,8462852 ordinated by increasing dates from top to bottom. The sky spectrum obtained simultaneously in fiber B has been fitted out of the original spectra (see Fig.~\ref{fig:NaI_spectrum}).}
\end{figure}

\begin{figure}\centering
\includegraphics[width=90mm, clip=true]{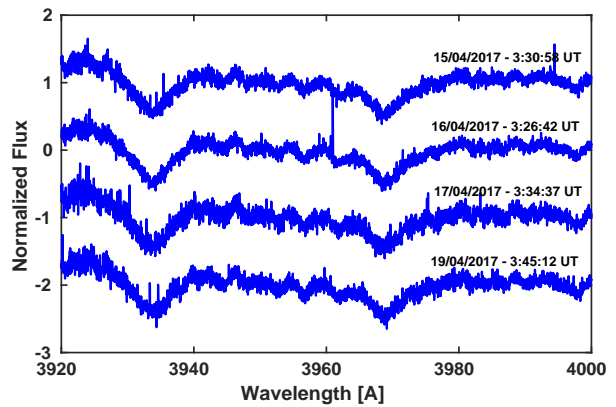}
\caption{\label{fig:CaII_spectrum_compared} Comparison of the Ca\,II spectra of KIC\,8462852 ordinated by increasing dates from top to bottom, with 3-pixels binning. As can be seen, there are no spectral signatures of transient phenomenon in these spectra.}
\end{figure}

Unfortunately, bad weather conditions prevented us to observe KIC\,8462852 after the 19th of March 2017. We could collect good quality spectra of KIC\,8462852 on the 15th, 16th, 17th and 19th of March between 03:30 and 03:45 UT. The median Na\,I spectrum of KIC\,8462852 observed with SOPHIE between these 4 dates is plotted on Fig.~\ref{fig:NaI_spectrum}. At the right hand side of the stellar Na\,I doublet lines, we detect an emission feature, which is also observed in the simultaneous sky-background spectrum obtained through the second aperture of the spectrograph (Fiber B). It is identified to geocoronal emission from Earth atmosphere. We subtracted this feature from all Na\,I spectra by fitting out the sky spectrum. As can be seen in Fig~\ref{fig:NaI_spectrum_compared}, the resulting Na\,I spectrum is totally quiet through the 4 days. It only presents a stable double peak absorption line, which is most likely of interstellar origin, since no counterpart is observed in the Ca\,II spectrum at the star radial velocity (Fig.~\ref{fig:CaII_spectrum_compared}). Similarly the Ca\,II doublet spectrum of KIC\,8462852 does not present any variable features between the 15th and the 19th of March.

Nevertheless, the predicted time of ingress is just after the observation dates. The 19th of March at 03:45 (UT) is at the top of the signature left wing, before the predicted timing of ingress (19th of March 04:48 UT). Therefore, the absence of observed features cannot exclude that significant absorption occurred in KIC\,8462852 spectrum during the transit. The observed spectra could neither infirm or confirm the periodicity of these transit events.

Assuming periodicity, the next event is predicted to occur between the 3rd and the 8th of October 2019 with ingress, centroid and egress timings given in Table~\ref{tab:timing_events}. New observations of KIC\,8462852 between the 3rd and the 8th of October 2019 in both photometry and spectroscopy, with Spitzer, Cheops, HST, JWST and ground-based spectroscopes, are strongly encouraged. They should allow confirmation or infirmation of the periodicity in the observed photometric event. 



\section{Conclusions}

After a careful detrending of the Kepler lightcurve of the peculiar star KIC\,8462852, we identified among 22 signatures, two strickingly similar shallow absorptions with a separation of 928.25 days (event A \& B). These two events presented 0.1\% deep stellar flux variations with duration of 4.4 days, consistent with the transit of a single or a few objects with a 928-days orbital period.

We thoroughly verified the different possible sources of systematics that could have produced the transit-like signatures of event A and B. We conclude that these two events are certainly of astrophysical origin, and occurred in the system of KIC\,8462852.

We found that two scenarios could well reproduce the transit lightcurve of events A and B. They consist in the occultation of the star by two kind of objects:

\begin{itemize}
\item[1. ] Either a string of half-a-dozen of exocomets orbiting at a distance $\gtrsim$0.3\,au, with evaporation rates similar to comet Hale-Bopp, and scattered along their common orbit much like the 1994 Shoemaker-Levy 9 fragments.
\item[2. ] Either an extended ring system surrounding a planet orbiting at 2.1\,au from the star, and composed of a constant opacity interior ring and an exterior ring with decreasing opacity towards larger radius.
\end{itemize}

It should be mentioned that the main argument against the exocomet scenario for KIC\,8462852 dimming events is the absence of any detectable IR excess. This is an important problem that will always lead to risky comparison with other emblematic exocomet hosts such as $\beta$\,Pic. These stars are all young ($<$100\,Myr) with strong Vega-like excess, and thus massive debris disk. The age of KIC\,8462852 (1\,Gyr) would well explain the lack of IR excess, still it must be explained how the vaporization of the remaining small bodies would fit below the detection level. In fact, \cite{Boyajian2016} showed that dust clouds of the mass of a fully vaporized Hale-Bopp comets, as needed to explain the strongest dips of KIC\,8462852 lightcurve, are not expected to produce visible IR emission as long as the distance of the clouds is greater than 0.2\,au. This happens to be the case in the exocomets string model proposed here.

This is the first strong evidence for a periodic signal coming from KIC\,8462852. All the other dimmings present irregular behavior with apparently uncorrelated timings. If periodic, our discovery opens a gate for in-depth characterization of a collection of objects present around this star. Assuming periodicity, we predict that the next event to happen will occur between the 3$^\text{rd}$ and the 8$^\text{th}$ of October 2019. The observation of KIC\,8462852 at these dates will confirm or infirm the 928.25\,day period, and hopefully will allow us to discriminate between the two scenarios proposed in this paper.

\begin{acknowledgements}
This work has been supported by the Centre National des Etudes Spatiales (CNES). We acknowledge the support of the French Agence Nationale de la Recherche (ANR), under program ANR-12-BS05-0012 “Exo-Atmos”. V.B. acknowledges the financial support of the Swiss National Science Foundation. This paper includes data collected by the Kepler NASA mission, and with the SOPHIE spectrograph on the 1.93-m telescope at Observatoire de Haute-Provence (CNRS). Funding for the Kepler mission is provided by the NASA Science Mission directorate. We thank the anonymous referee for his careful reading of our manuscript and his insightful comments and suggestions.
\end{acknowledgements}

%
%

\end{document}